\documentclass[aps,pre,reprint,superscriptaddress,amsmath,showpacs]{revtex4-1}
\pdfoutput = 1
\usepackage{graphicx}
\usepackage{amssymb}
\usepackage{sidecap}
\usepackage{color}
\definecolor{darkblue}{rgb}{0,0,0.6}
\definecolor{darkred}{rgb}{0.6,0,0}
\definecolor{darkgreen}{rgb}{0,0.6,0}
\usepackage[colorlinks=true,urlcolor=darkblue,citecolor=darkblue,linkcolor=darkred,hyperfootnotes=false]{hyperref}

\begin{document}

\title{Effects of a local defect on one-dimensional nonlinear surface growth}

\author{Hyungjoon Soh}
\affiliation{Department of Physics,
Korea Advanced Institute of Science and Technology, Daejeon
34141, Korea}

\author{Yongjoo Baek}
\affiliation{Department of Physics, Technion--Israel Institute of Technology, Haifa 32000, Israel}

\author{Meesoon Ha}
\email[Corresponding author: ]{msha@chosun.ac.kr}
\affiliation{Department of Physics Education, Chosun University,
Gwangju 61452, Korea}

\author{Hawoong Jeong}
%\email[ ]{hjeong@kaist.edu}
\affiliation{Department of Physics and Institute for the
BioCentury, Korea Advanced Institute of Science and Technology,
Daejeon 34141, Korea}

\date{\today}

\begin{abstract}
The slow-bond problem is a long-standing question about the minimal strength $\epsilon_\mathrm{c}$ of a local defect with global effects on the Kardar--Parisi--Zhang (KPZ) universality class. A consensus on the issue has been delayed due to the discrepancy between various analytical predictions claiming $\epsilon_\mathrm{c} = 0$ and numerical observations claiming $\epsilon_\mathrm{c} > 0$. We revisit the problem via finite-size scaling analyses of the slow-bond effects, which are tested for different boundary conditions through extensive Monte Carlo simulations. Our results provide evidence that the previously reported nonzero $\epsilon_\mathrm{c}$ is an artifact of a crossover phenomenon, which logarithmically converges to zero as the system size goes to infinity.
\end{abstract}


%----------------------------------------------------------------------------
%02.50.-r Probability theory, stochastic processes, and statistics
%05.40.-a Fluctuation phenomena, random processes, noise, and Brownian motion
%05.70.Ln Nonequilibrium and irreversible thermodynamics
%64.60.-i General studies of phase transitions
%64.60.Ht Dynamic critical phenomena
%----------------------------------------------------------------------------

\maketitle

\section{Introduction} 
\label{sec:intro}

The Kardar--Parisi--Zhang (KPZ) equation~\cite{Kardar1986} for the one-dimensional (1D) case describes a broad range of nonequilibrium phenomena, such as nonlinear surface growth~\cite{Barabasi1995,HalpinHealy1995,Krug1997}, biased transport of particles~\cite{Derrida1992,*Schutz1993,*Blythe2007}, fluctuations of directed polymers~\cite{Kardar1987,HalpinHealy1995}, and so on. Being a rare example of nonequilibrium processes whose steady-state statistics are fully known, the KPZ equation has received much attention as the governing dynamics underlying a universality class beyond equilibrium~\cite{Corwin2011,Quastel2015}. More recently, properties of the KPZ universality class have also been studied via concrete experimental realizations~\cite{Takeuchi2010,*Takeuchi2011,*Takeuchi2013,*DeNardis2017}.

In spite of these advances, open questions still remain. One important example is whether a {\em local} defect is relevant to macroscopic properties. Nonequilibrium systems with a conserved quantity generically display weak long-range correlations, even far away from criticality~\cite{Spohn1983,Garrido1990,Dorfman1994}. This opens up the possibility that even local perturbations may have global consequences in the presence of nonequilibrium driving; however, there is no known general criterion for determining how strong a local defect should be to induce such phenomena. For 1D systems governed by the KPZ equation, this issue gave rise to the {\em slow-bond} (SB) problem, which has been a subject of much debate.

\begin{figure}[t]
%% Figure 1 : Schematic BCSOS growth and TASEP with a SB
	\centering
   \includegraphics[width=0.85\columnwidth]{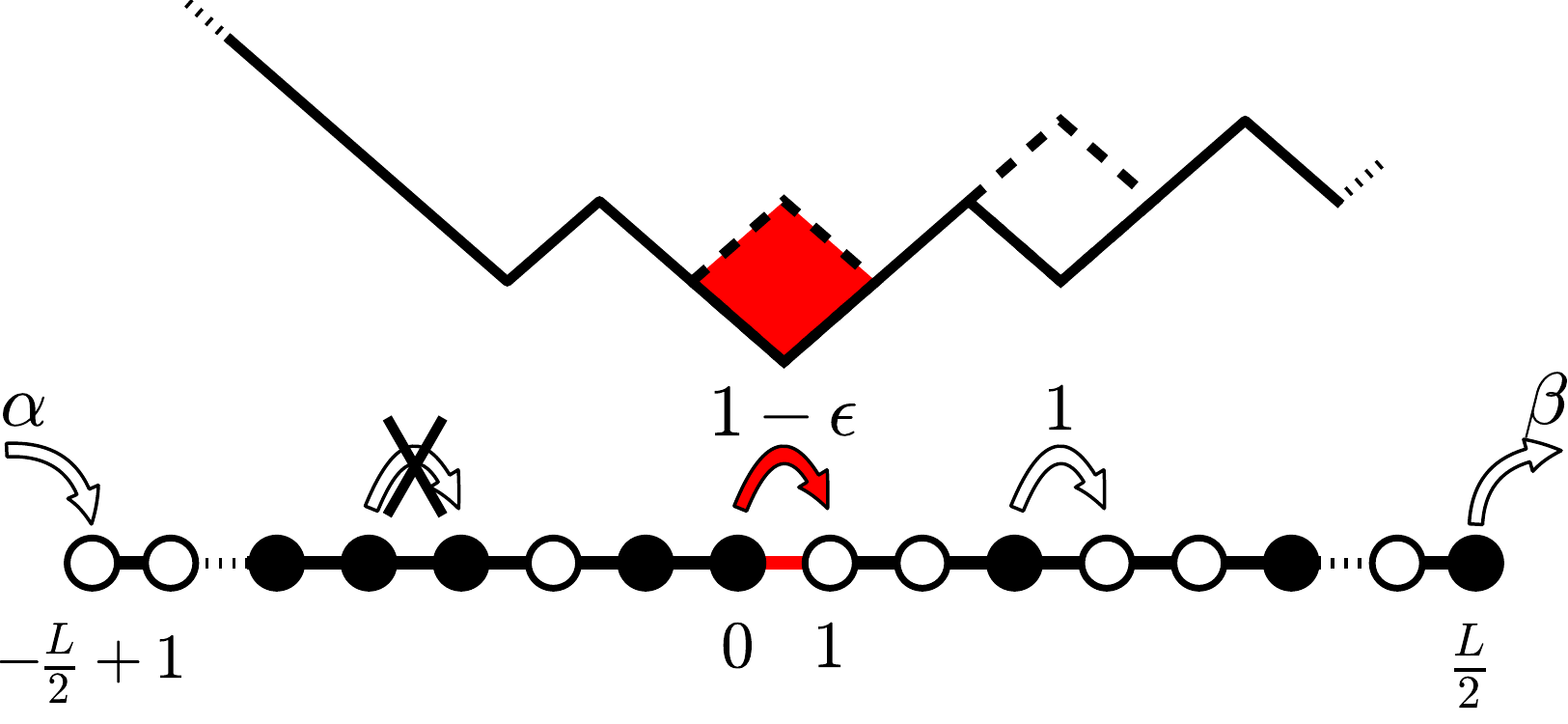}
	\caption{\label{fig:model} 1D TASEP (lower) and the corresponding BCSOS growth model (upper) with a slow bond of strength $\epsilon \in [0,1]$ (marked by a red segment) and open boundaries. The numbers above the arrows indicate hopping rates, and the site indices are shown at the bottom.}
\end{figure}

\begin{figure}[b]
%% Figure 2 : Schematic Phase Diagram	
 \centering
   \includegraphics[width=0.9\columnwidth]{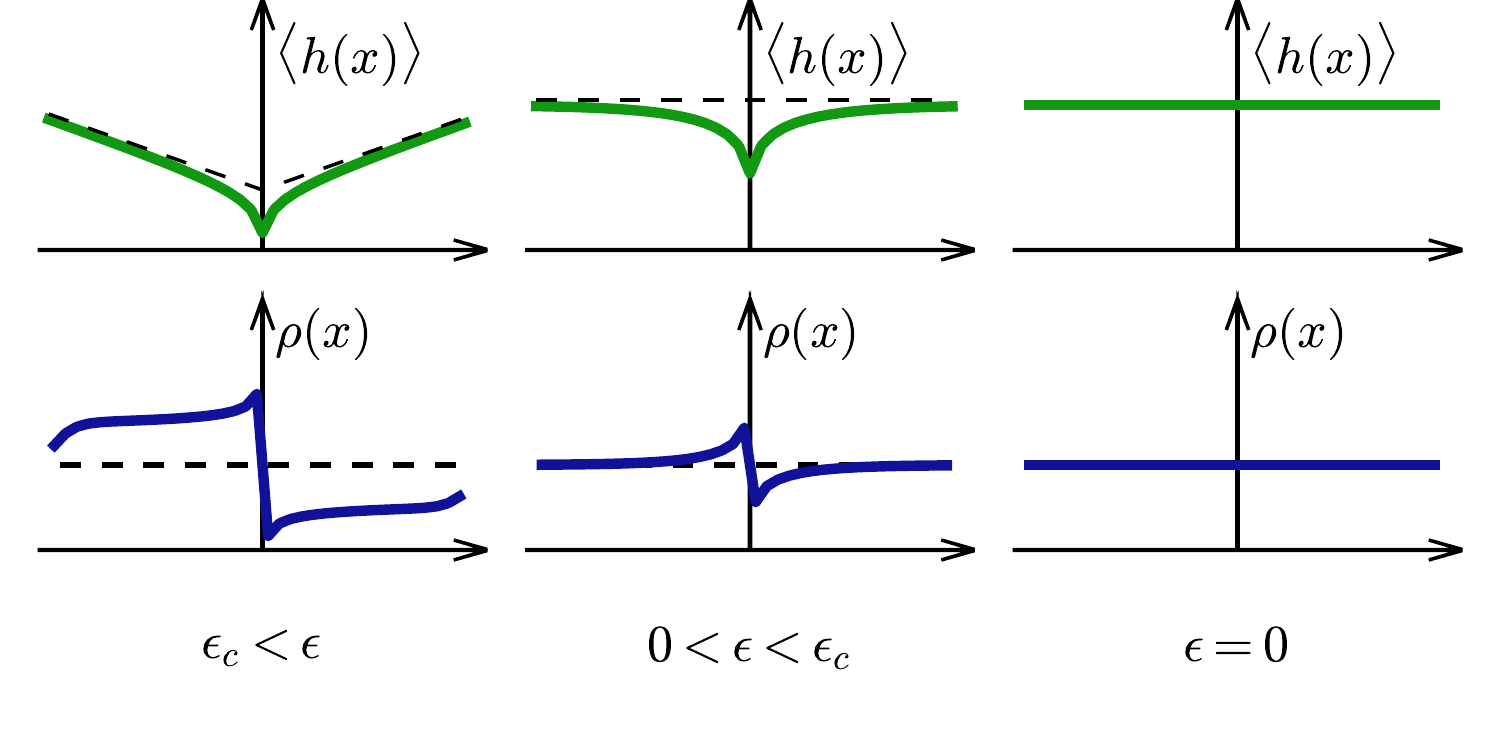}
	\caption{\label{fig:PD} Average height profiles of the BCSOS growth model (upper) and density profiles of the TASEP (lower) under the assumption of $\epsilon_\mathrm{c} > 0$. In the strong SB regime ($\epsilon > \epsilon_\mathrm{c}$), the TASEP (BCSOS growth model) develops a queue (nonzero slope) spanning the system. In the weak SB regime ($0 < \epsilon < \epsilon_\mathrm{c}$), the queue (nonzero slope) disappears on a large scale, leaving algebraically decaying corrections around the slow bond. In the absence of the slow bond ($\epsilon = 0$), the queue (nonzero slope) vanishes.}
\end{figure}

Using the language of lattice gases, the SB problem can be stated as follows~\cite{*[{The SB problem can also be posed in the context of condensation transitions; See~}] [{}] Cirillo2016}. Consider self-driven particles with excluded volume interactions hopping in one direction through a 1D lattice. If the hopping is slowed down by a factor of $1-\epsilon$ across one fixed bond (see Fig.~\ref{fig:model}), how {\em slow} should the bond be to create a queue of particles before it (or vacancies after it) on a macroscopic length scale? Two conflicting hypotheses have been proposed. On one side, a slow bond of any strength is claimed to have long-range effects (i.e., the critical strength is $\epsilon_\mathrm{c} = 0$). This claim is supported by approximate analytical arguments based on the mean-field theory~\cite{Janowsky1992,Kolomeisky1998a}, series expansions with respect to small $\epsilon$~\cite{Janowsky1994,Costin2012}, and renormalization group arguments~\cite{Tang1993,Balents1994,*Kinzelbach1995,*Hwa1995,*Kolomeisky1995,*Lassig1998}. On the other side, a {\em nonqueued SB phase} is claimed to exist (for $0 < \epsilon < \epsilon_\mathrm{c}$), in which the effects of the slow bond decay algebraically to zero with distance, with no large-scale queue (see Fig.~\ref{fig:PD}). This hypothesis was put forth by \cite{MHa2003} based on finite-size scaling (FSS) of numerical results, with an estimate of $\epsilon_\mathrm{c} = 0.20(2)$. Similar claims for $\epsilon_\mathrm{c} > 0$ can also be found in other numerical studies~\cite{Kandel1992,JHLee2009} and notably in an experiment on the slow combustion of a paper with a columnar defect~\cite{Myllys2003}. Due to the lack of exact analytical methods, the controversy has remained unresolved for more than a couple of decades~\cite{*[{We note that, when the updates are parallel, the TASEP with a slow bond was analytically solved, yielding $\epsilon_\mathrm{c} = 0$}; See~] [{}] SchutzJSP1993,*SchutzPRE1993,*Scoppola2015.}.

More recently, Basu {\it et al.}\cite{Basu2014} claimed to have rigorously proven $\epsilon_\mathrm{c} = 0$. An ensuing report on high-precision simulations~\cite{Schmidt2015} seems to indicate that $\epsilon_\mathrm{c}$ is much smaller than previously claimed, with hallmarks of global SB effects persistently detected down to $\epsilon \ge 0.01$. These results call for a reconsideration of the previous studies claiming $\epsilon_\mathrm{c} > 0$. In particular, there arises a question whether the FSS hypothesis of \cite{MHa2003} can be replaced with a new one which is consistent with $\epsilon_\mathrm{c} = 0$. Furthermore, there are some subtle but important differences in the boundary conditions employed in different numerical studies (see Sec.~\ref{ssec:problem} for details). In order to address the seeming disagreement between the studies more thoroughly, we should check whether consistent observations can be made irrespective of the boundary conditions.

In this paper, we revisit the SB problem via extensive Monte Carlo simulations for both open and closed boundary conditions. We propose a FSS hypothesis for the steady-state current, which is consistent with an essential singularity at $\epsilon_\mathrm{c} = 0$ suggested by \cite{Janowsky1994,Costin2012}. In this viewpoint, the nonqueued SB phase proposed in \cite{MHa2003} is not a true dynamical phase but only a crossover phenomenon caused by the system size being less than the correlation length.

The rest of the paper is organized as follows. In Sec.~\ref{sec:model}, we first introduce the SB problem and its representations in two simplest examples of the KPZ universality class, namely the totally asymmetric simple exclusion process (TASEP) and the body-centered solid-on-solid (BCSOS) model of surface growth. In Sec.~\ref{sec:current}, we describe the behaviors of the steady-state current obtained from extensive Monte Carlo simulations. FSS analyses of the relevant observables are presented in Sec.~\ref{sec:fss}. Finally, a summary of our results is given in Sec.~\ref{sec:summary}.

%%%Maybe invoke Ensemble Equivalence here
%%%%

\section{Models and problems}
\label{sec:model}

\subsection{TASEP with a slow bond} \label{ssec:tasep}

One of the simplest lattice gas models belonging to the KPZ universality class is the totally asymmetric simple exclusion process (TASEP). Here we present a standard discrete-time implementation of the model in the presence of a slow bond.

We consider a 1D lattice with $L$ sites. For convenience, $L$ is assumed to be an even number, so that the sites can be indexed by $x = -L/2+1,\ldots,0,\ldots,L/2$. Each site can be occupied by at most a single particle, so that the occupancy of site $x$ satisfies $n(x) \in \left\{0, 1\right\}$. At each time step, a randomly chosen bond is updated by
\begin{align}
	\bullet\,\circ \longrightarrow \circ\,\bullet
\end{align}
if and only if the left end of the bond is occupied ($\bullet$) and the right end is vacant ($\circ$). The only exception to this rule is that, for the bond between sites $0$ and $1$, the above update takes place with probability $1-\epsilon$ with $\epsilon \ge 0$. Thus the bond between sites $0$ and $1$ is our slow bond.

There are two possible boundary conditions for the system. For the periodic boundary condition (PBC), the leftmost site $-L/2+1$ is adjacent to the rightmost site $L/2$, so that the system forms a closed loop. Since the number of particles is fixed (say at $N \le L$), the density of particles $\rho_0 \equiv N/L$ can be regarded as the control parameter. On the other hand, for the open boundary condition (OBC), each end of the system is connected by a bond to its own particle reservoir (see Fig.~\ref{fig:model}). If the bond between the left (right) reservoir and site $-L/2+1$ ($L/2$) is chosen, provided that the site is vacant (occupied), a particle jumps into (out of) the site with probability $\alpha$ ($\beta$). Note that for the OBC the slow bond is exactly in the middle of the system. For reasons to be explained later (see Sec.~\ref{ssec:problem}), we choose $\rho_0 = 1/2$ for the PBC and $\alpha = \beta = 1/2$ for the OBC.

In order to implement the continuous flow of time, after each step the time is increased by
\begin{align} \label{eq:t_upd_ord}
	t \longrightarrow t + \frac{\ln (1/u)}{\text{number of bonds}},	
\end{align}
where $u \in [0,1)$ is a uniformly distributed random number. This ensures that each bond is chosen by a Poisson process of rate $1$. Since the fluctuations of the Poisson process are irrelevant to the observed steady-state statistics, we can reduce the computational cost without affecting the results by replacing $\ln (1/u)$ with $1$, so that the time is updated deterministically. Even then the number of bonds in the denominator sets the correct time scale for our simulation.

To make our discussions self-contained, we give a brief account of the dynamical phases of the TASEP in the limit $L \to \infty$ for $\epsilon = 0$, which can be described by both mean-field approximations and exact methods~\cite{Derrida1992,*Schutz1993,*Blythe2007}. In the steady state, the system develops a bulk region in which the average density of particles forms a flat profile $\rho(x) \equiv \langle n(x) \rangle = \rho_\mathrm{b}$, where $\langle \cdot \rangle$ denotes an ensemble average. Since correlations between adjacent sites become zero in this region, the steady-state current $J$, defined as the average number of particles hopping through each bond per unit time, is related to the bulk density $\rho_\mathrm{b}$ by
\begin{align} \label{eq:J_rb}
	J &= \left\langle n(x)\left[1-n(x+1)\right] \right\rangle = \rho(x)\left[1-\rho(x+1)\right] \nonumber\\
	&= \rho_\mathrm{b}(1-\rho_\mathrm{b}).
\end{align}

For the PBC, we always have a flat average density profile throughout the system, so $\rho_\mathrm{b} = \rho_0$ and $J = \rho_0 (1-\rho_0)$. In this case the maximal-current phase, characterized by $J = 1/4$ and $\rho_\mathrm{b} = 1/2$, is attained only at $\rho_0 = 1/2$; otherwise, for $\rho_0 < 1/2$ ($\rho_0 > 1/2$) the system is in the low-density (high-density) phase characterized by excess hole (particle) density in the bulk.

On the other hand, for the OBC the bulk density is given by
\begin{align}
	\rho_\mathrm{b} = \begin{cases}
	\min\left[\alpha,1/2\right] &\text{if $\alpha \le \beta$;} \\
	\max\left[1-\beta,1/2\right] &\text{otherwise.}
 \end{cases}
\end{align}
In this case the maximal current $J = 1/4$ is attained only for $\alpha \ge 1/2$ and  $\beta \ge 1/2$; otherwise, the system is in the low-density (high-density) phase for $\alpha < \beta$ ($\alpha > \beta$), in which the bulk density $\rho_\mathrm{b}$ is controlled by the entry rate $\alpha$ (exit rate $\beta$). If $\alpha = \beta < 1/2$, the system exhibits a phase separation into a pair of low-density and high-density regions, with a sharp interface (called {\em shock}) formed between the two.

\subsection{Mapping to surface growth} \label{ssec:bcsos}

It is often useful to reinterpret the TASEP in terms of a related surface growth process called the body-centered solid-on-solid (BCSOS) growth model~\cite{vanBeijeren1977}. The mapping from a particle configuration of the TASEP to a height profile of the BCSOS growth model is simply given by the relation
\begin{align}
	h(x) \equiv h\left(-\frac{L-1}{2}\right) + \sum_{x'=-\frac{L}{2}+1}^{x-\frac{1}{2}} \left[1 - 2n(x')\right]
\end{align}
for $x = -\frac{L-3}{2}, -\frac{L-5}{2}, \ldots, \frac{L+1}{2}$ (see Fig.~\ref{fig:model}). This implies that the height difference between adjacent sites is always $\Delta h = \pm 1$ and that a hopping from site $x$ to site $x+1$ corresponds to the deposition $h(x+1/2) \to h(x+1/2) + 2$. (See the tilted blocks with dashed boundaries in Fig.~\ref{fig:model}.) Similarly, each entry from the left reservoir (exit to the right reservoir) corresponds to an increase of the height at $x = -\frac{L-1}{2}$ ($x = \frac{L+1}{2}$) by 2. Due to the slow bond, the deposition also slows down by a factor of $1-\epsilon$ at $x = \frac{1}{2}$.

\subsection{Slow-bond (SB) problem} \label{ssec:problem}

With two simple implementations of the KPZ universality class introduced above, we now give a precise formulation of the SB problem in terms of these models. We consider the case when the system is, if there is no slow bond ($\epsilon = 0$), in the maximal-current phase ($J = 1/4$, $\rho_\mathrm{b} = 1/2$). This means that the control parameters should be given by $\rho_0 = 1/2$ for the PBC and $\alpha \ge 1/2$, $\beta \ge 1/2$ for the OBC. We note that $\alpha = \beta = 1$ was used in \cite{Schmidt2015}, which puts the system deep inside the maximal-current regime to avoid possible complications near the phase boundary. On the other hand, $\alpha = \beta = 1/2$ was used in \cite{MHa2003}, with the purpose of flattening the density profile near the the reservoirs when the bulk density satisfies $\rho_\mathrm{b} = 1/2$. This ensures that, in the weak SB regime, the absence or presence of a queue becomes easier to detect. In this study, we use the latter condition for a closer comparison with \cite{MHa2003}. Once the control parameters are thus fixed, the SB strength $\epsilon$ becomes the only free parameter.

The critical strength $\epsilon_\mathrm{c}$ of the slow bond is defined as follows. If $0 \le \epsilon \le \epsilon_\mathrm{c}$, the steady-state current remains at its maximal value $J = 1/4$. Equivalently, the bulk density before and after the slow bond is given by its maximal-current value $\rho_\mathrm{b} = 1/2$. In the language of the BCSOS growth model, the average height profile $\langle h(x) \rangle$ remains largely flat before and after the slow bond.

On the other hand, if $\epsilon > \epsilon_\mathrm{c}$, $J$ becomes smaller than $1/4$, which is equivalent to an increase (decrease) of $\rho_\mathrm{b}$ from $1/2$ before (after) the slow bond. We may denote the change of the bulk density by $\Delta_\mathrm{b}(\epsilon) \equiv \rho_\mathrm{b}(\epsilon) - 1/2$ and use it as an indicator of whether a macroscopic queue is formed around the slow bond. We note that $-\Delta_\mathrm{b}$ can also be interpreted as the average slope of the height profile in the BCSOS growth model. 

The SB problem is a question of the precise value of $\epsilon_\mathrm{c}$. In Fig.~\ref{fig:PD}, under the assumption that $\epsilon_\mathrm{c} > 0$ according to the scenario proposed by \cite{MHa2003}, we give schematic illustrations of the height profiles of the BCSOS growth model (upper panel) and the density profiles of the TASEP (lower panel). Boundaries at the slow bond, reservoirs (for the OBC), or between high-density and low-density regions (for the PBC) generally produce some nonlinear modulations, which gradually decay as one moves away from the boundaries into the bulk. If the nonqueued SB phase $0 < \epsilon < \epsilon_\mathrm{c}$ exists, it can be distinguished from the case $\epsilon = 0$ by such density (or height) modulations near the slow bond (see the middle column of Fig.~\ref{fig:PD}). It was claimed in \cite{MHa2003} that these boundary effects show a power-law decay. On the other hand, if $\epsilon_\mathrm{c} = 0$, the nonqueued SB phase would simply be nonexistent.

Based on these scenarios, we aim to estimate the value of $\epsilon_\mathrm{c}$ for both PBC and OBC by extensive Monte Carlo simulations.

%%%%%%%
\begin{figure*}[t]
%% Figure: Steady-state current
	\centering
	\includegraphics[width=\textwidth]{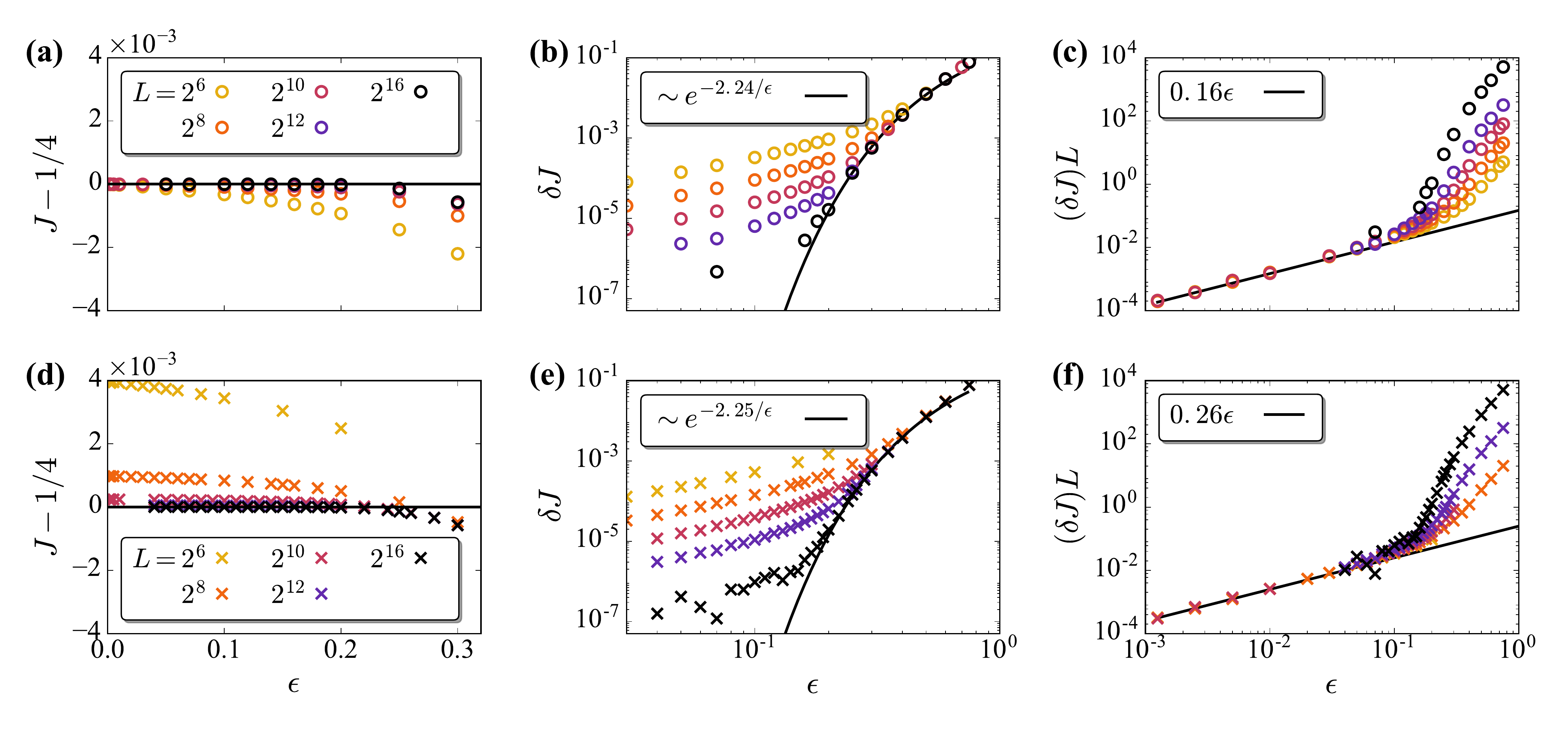}
        \caption{\label{fig:current} The steady-state current $J$ and the reduced current $\delta J$ as the SB strength $\epsilon$ and the system size $L$ are varied from $L = 2^6$ to $2^{16}$, with darker colors used for larger $L$. (a) For the OBC ($\circ$), $J$ monotonically converges to $1/4$ (black solid line) as $\epsilon$ goes to $0$, regardless of $L$. (b) For large $\epsilon$, we observe $\delta J \sim \exp(-b/\epsilon)$ (black solid line), with $b \simeq 2.24$ estimated by curve fitting. (c) For small $\epsilon$, we observe $\delta J \simeq \epsilon/L$ (black solid line). (d--f) Similar plots are shown for the PBC ($\times$).}
\end{figure*}

\subsection{Simulation method} \label{ssec:method}

For any boundary condition, we start each run of the simulation from the initial condition in which one particle is occupying every second site (``$\bullet\,\circ\,\bullet\,\circ\,\cdots$''). For the PBC, this initial condition automatically sets $\rho_0 = 1/2$. As long as $\rho_0$ is correctly set (for the PBC) and the simulation time is sufficiently longer than the relaxation time, the precise form of the initial condition is irrelevant to the results. Assuming that the decay time for any correlation in the system grows as $L^{3/2}$ (KPZ scaling) or slower, we let the system evolve until $t = 100L^{3/2}$ in order to ensure that the system reaches the steady state. After then we measure the observables of interest at every $L^{3/2}$ time steps, so that all measurements can be regarded as independent of each other.

In order to reduce the simulation time, we keep a list of ``active bonds'', which are bonds with the right configuration (``$\bullet\,\circ$'') for a hopping to occur. The bond connecting the leftmost (rightmost) site to the adjacent reservoir is also included in the list if the site is vacant (occupied). At each step, one of the bonds in the list is chosen at random and updated according to the rules of the process. This method decreases the number of random number generations~\footnote{Here the Mersenne Twister is used as a pseudo-random number generator, which is one of the most widely used. See \url{http://www.math.sci.hiroshima-u.ac.jp/~m-mat/MT/ARTICLES/earticles.html}} required for each hopping to occur. Instead, each update should increase the time by
\begin{align} \label{eq:t_upd_rf}
	t \longrightarrow t + \frac{\ln (1/u)}{\text{number of active bonds}},	
\end{align}
where $u \in [0,1)$ is a uniformly distributed random number. As previously discussed, $\ln (1/u)$ can be replaced with $1$ for convenience. This scheme increases the time faster than Eq.~\eqref{eq:t_upd_ord} does.

\section{Steady-state current} \label{sec:current}

As explained in Sec.~\ref{ssec:problem}, the steady-state current $J$ is a direct indicator of the SB effects. In this section, we present numerical results showing how $J = J(\epsilon;L)$ depends on the SB strength $\epsilon$ and the system size $L$. We first note that, in the absence of a slow bond ($\epsilon = 0$), the value of $J$ is exactly known for any value of $L$. For the OBC with $\alpha = \beta = 1/2$, due to the exact particle--hole symmetry, we have $J(0;L) = 1/4$ for any value of $L$, as shown in Fig.~\ref{fig:current}(a). For the PBC, due to nonvanishing correlations at finite size, we have $J(0;L) = \frac{1}{4(1-1/L)}$~\cite{Derrida1992,*Schutz1993,*Blythe2007}. This is indicated by $J > 1/4$ observed for small values of $\epsilon$ in Fig.~\ref{fig:current}(d). With these properties in mind, we choose the reduced current
\begin{align}
	\delta J(\epsilon;L) \equiv J(0;L) - J(\epsilon;L)
\end{align}
to be the indicator of the SB effect for each system size $L$.

\subsection{Case of large $\epsilon$}

Some analytical studies~\cite{Janowsky1994,Costin2012} argued that, in the limit $L \to \infty$, $\delta J$ obeys the relation
\begin{align} \label{eq:DJ-exp-fit}
	\delta J \sim \exp(-b/\epsilon),
\end{align}
where $b \simeq 2$ according to \cite{Costin2012}. If true, this relation supports $\epsilon_\mathrm{c} = 0$ and confirms the existence of an essential singularity at the transition point, which was predicted by renormalization group arguments~\cite{Tang1993,Balents1994,*Kinzelbach1995,*Hwa1995,*Kolomeisky1995,*Lassig1998}. In what follows we numerically check the plausibility of Eq.~\eqref{eq:DJ-exp-fit}.

Figure~\ref{fig:current}(b) shows the behaviors of $\delta J(\epsilon,L)$ for the OBC with $\alpha = \beta = 1/2$. For large $\epsilon$, $\delta J$ closely follows the exponential form shown in Eq.~\eqref{eq:DJ-exp-fit}, as shown by the agreement between the data points in this regime and the black solid curve corresponding to $b \simeq 2.25$ obtained from the method of least squares (applied to data points with $0.2 < \epsilon < 0.5$). This estimated value of $b$ is slightly higher than $b \simeq 2$ suggested by \cite{Costin2012}. Since Eq.~\eqref{eq:DJ-exp-fit} describes the leading-order behavior in the limit $\epsilon \to 0$, this deviation might be a systematic bias due to higher-order contributions in $\epsilon$.

Similar observations can be made for the PBC, as shown in Fig.~\ref{fig:current}(e). For large $\epsilon$, $\delta J$ is again well described by Eq.~\eqref{eq:DJ-exp-fit} with the least-squares estimate $b \simeq 2.24$, which is very close to the corresponding value for the PBC.

\subsection{Case of small $\epsilon$}

As shown in Figs.~\ref{fig:current}(b) and \ref{fig:current}(e), for small $\epsilon$, the agreement with Eq.~\eqref{eq:DJ-exp-fit} becomes poor. In \cite{MHa2003}, this lack of agreement was interpreted as numerical evidence against Eq.~\eqref{eq:DJ-exp-fit}. However, Figs.~\ref{fig:current}(b) and \ref{fig:current}(e) also indicate that $\delta J$ has a strong dependence on $L$ for small $\epsilon$, which implies the significance of finite-size effects in this regime. In order to check the consistency between Eq.~\eqref{eq:DJ-exp-fit} and the observed small-$\epsilon$ behaviors of $\delta J$, we must identify those finite-size effects.

Figures~\ref{fig:current}(c) and \ref{fig:current}(f) show that, for both OBC and PBC, $\delta J$ shows a linear scaling behavior
\begin{align} \label{eq:DJ-lin-fit}
	\delta J \sim \epsilon/L
\end{align}
for extremely small values of $\epsilon$. When $\epsilon$ is much smaller than the inverse of the relaxation time scale $\tau \sim L^{3/2}$, each blocking event at the slow bond becomes independent of the other blocking events. Then the rate of blocking events becomes proportional to $\epsilon$ and independent of $L$. Since the reduced current $\delta J$ is obtained by dividing the rate of blocking events by $L$, the scaling behavior of Eq.~\eqref{eq:DJ-lin-fit} naturally arises in this regime.

To complete our picture, we need to know how the exponential behavior of Eq.~\eqref{eq:DJ-exp-fit} changes to different scaling behaviors for smaller $\epsilon$, whose lowest extreme is the linear behavior of Eq.~\eqref{eq:DJ-lin-fit}. If $\epsilon_\mathrm{c} = 0$, the crossover should occur at a size-dependent threshold $\epsilon_\times(L)$, which approaches zero as $L$ goes to infinity. More details about this crossover phenomenon are given in the next section.

\begin{figure}[b]
	\centering 
	\includegraphics[width=\columnwidth]{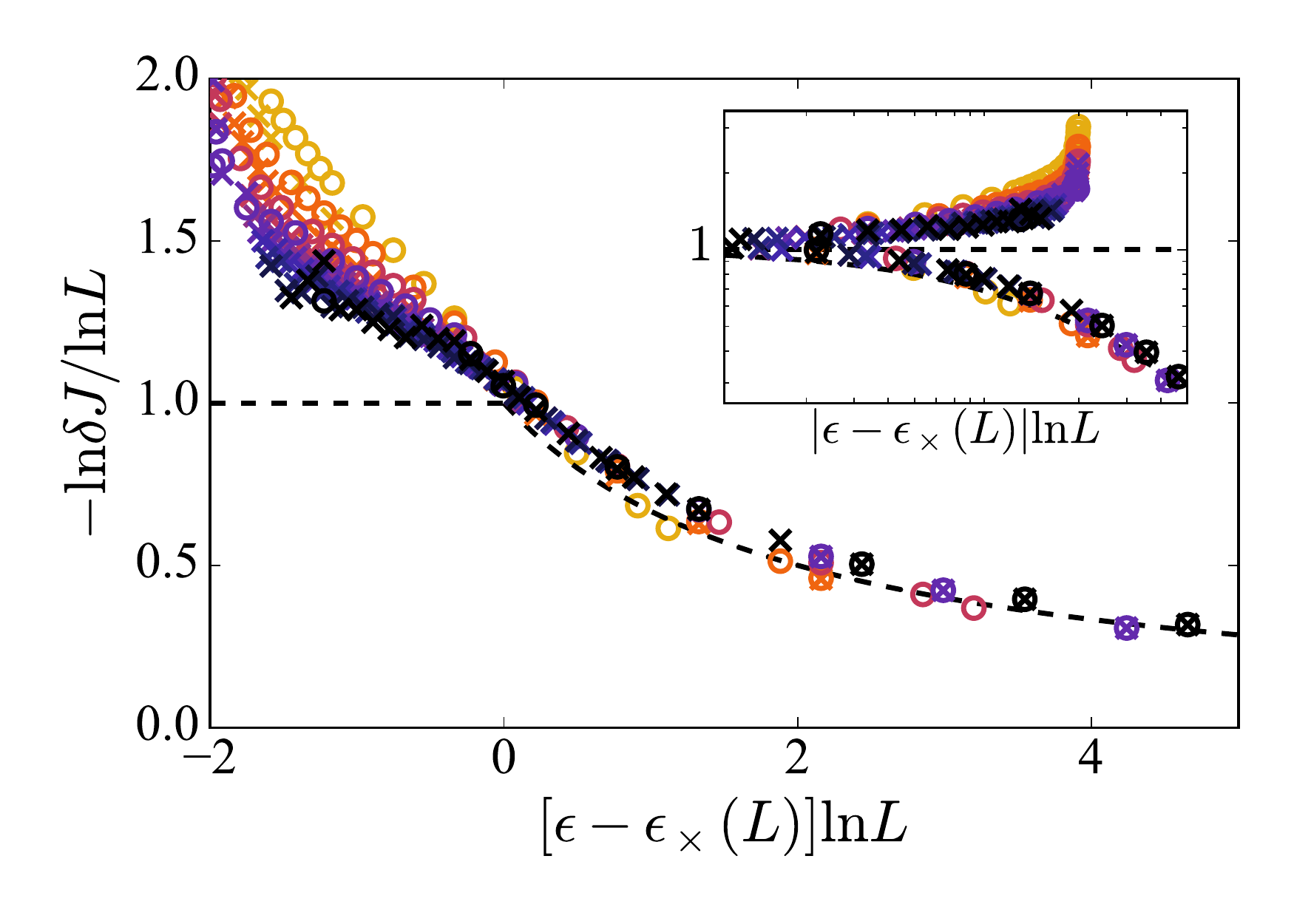}
	\caption{\label{fig:fss} Data collapse by Eq.~\eqref{eq:DJ_fss}, where the crossover scale is given by $\epsilon_\times(L) = b/\ln L$. Using the estimates of $b$ shown in Fig.~\ref{fig:current}, the data for the OBC ($\circ$) and the PBC ($\times$) are compared with the asymptotic scaling function $\Psi_{J}(x)$ (dashed black lines) given by Eq.~\eqref{eq:psi_DJ_def}. The system size is varied from $L = 2^6$ to $2^{16}$, with darker colors used for larger $L$ (see the legends of Fig.~\ref{fig:current}). The inset is a log-log replot of the main figure.}
\end{figure}

\section{Finite-size scaling (FSS)} \label{sec:fss}

\subsection{FSS form of $\delta J$}

Here we propose a FSS form which interpolates Eqs.~\eqref{eq:DJ-exp-fit} and \eqref{eq:DJ-lin-fit} in a manner consistent with $\epsilon_\mathrm{c} = 0$. The crossover SB strength $\epsilon_\times(L)$ can be obtained from the following simple argument. When $\epsilon > \epsilon_\times(L)$, a high-density (low-density) bulk region is formed before (after) the slow bond. According to the exact solution of the 1D TASEP with open boundaries~\cite{Derrida1992,Schutz1993}, the particle density near the entrance (exit) of the system in the high-density (low-density) phase algebraically approaches the bulk value $\rho_\mathrm{b}$, with an exponential cutoff characterized by the correlation length $\xi(J)$ given by
\begin{align} \label{eq:xi_DJ}
	\xi(J) \equiv -\frac{1}{\ln (4J)} \sim \frac{1}{\delta J}
\end{align}
for small $\delta J$. When the slow bond is weak, we may combine Eqs.~\eqref{eq:DJ-exp-fit} and \eqref{eq:xi_DJ} to obtain
\begin{align}
	\xi(J) \sim e^{b/\epsilon}.
\end{align}
As $\epsilon$ is further decreased, the system starts to be influenced by finite-size effects when $e^{b/\epsilon} \sim L$. This implies that the crossover scale $\epsilon_\times(L)$ is given by
\begin{align} \label{eq:epsilon_crossover}
	\epsilon_\times(L) = \frac{b}{\ln L}.
\end{align}

Now we construct a FSS theory which describes the behavior of $\delta J$ for $\epsilon\sim 1/\ln L$. We propose a FSS form
\begin{align} \label{eq:DJ_fss}
	\frac{\ln \delta J}{\ln L} = \Psi_{J}\left[|\epsilon-\epsilon_\times(L)|\ln L\right],
\end{align}
where
\begin{align} \label{eq:psi_DJ_def}
	\Psi_{J}(\zeta) =
	\begin{cases}
		-1 & \text{ for $-b < \zeta < 0$},\\
		-\frac{b}{x+b} & \text{ for $\zeta > 0$}.	
	\end{cases}
\end{align}
One can easily check that the behavior of $\Psi_J(\zeta)$ for $\zeta > 0$ is consistent with Eq.~\eqref{eq:DJ-exp-fit}. To obtain the behavior for $\zeta < 0$, we use Eq.~\eqref{eq:DJ-lin-fit} as well as the fact that $\delta J$ monotonically increases with $\epsilon$, which in turn implies that $\Psi_J(\zeta)$ should be a monotonic function. As shown in Fig.~\ref{fig:fss}, our numerical data are in good agreement with Eq.~\eqref{eq:DJ_fss}, thus supporting our scenario that $\epsilon_\mathrm{c} = 0$ is achieved by the crossover scale $\epsilon_\times(L) \sim 1/\ln L$ approaching zero as $L$ goes to infinity.

\begin{figure}[b]
	\centering
	\includegraphics[width=\columnwidth]{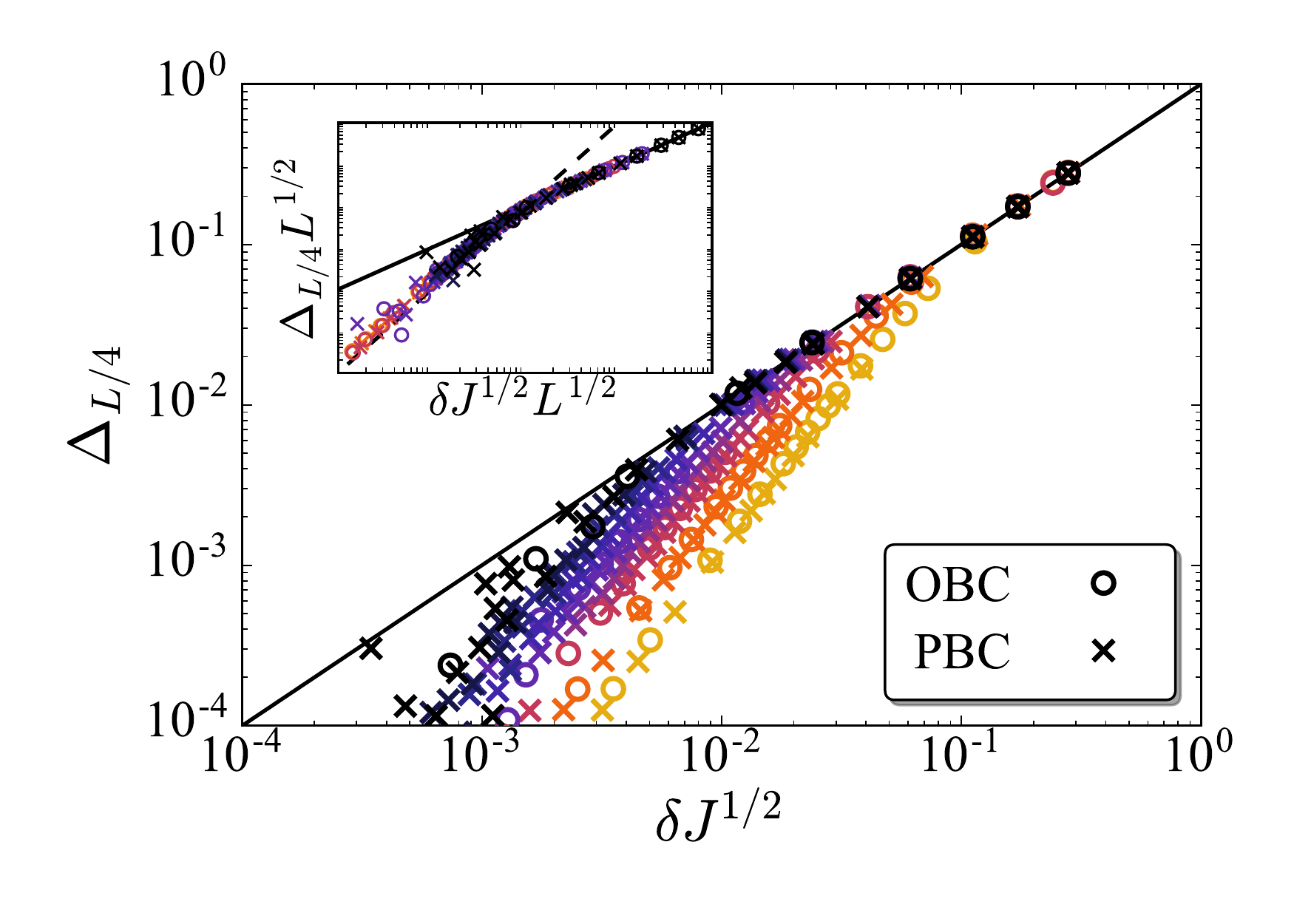}
	\caption{\label{fig:boundary_effects} Relations between the excess particle density $\Delta_{L/4}$ and the reduced current $\delta J$. The black diagonal line corresponds to $\Delta_{L/4} = \delta J$. The inset shows the data collapse by Eq.~\eqref{eq:DL4_fss}, with the black solid (dashed) line indicating $y = x$ ($y = x^2$). The system size ranges from $L = 2^6$ to $2^{16}$, with darker colors used for larger $L$  (see the legends of Fig.~\ref{fig:current}).}
\end{figure}

\subsection{Nature of the weak SB regime}

In order to get more insight into the weak SB regime bounded by $0 < \epsilon < \epsilon_\times(L)$, we study the excess particle density
\begin{align}
	\Delta_{L/4} \equiv |\rho_{L/4} - 1/2|
\end{align}
at the site $L/4$, which is exactly in the middle of a half-system before or after the slow bond. This observable is chosen as an indicator of the SB effects on the deep interior of the system.

Varying $\epsilon$ and $L$, we numerically observe the relations between $\Delta_{L/4}$ and $\delta J$ for the OBC and the PBC, which are plotted in Fig.~\ref{fig:boundary_effects}. The main plot shows that $\Delta_{L/4}$ stays close to $\delta J^{1/2}$ only for $\delta J$ larger than a size-dependent threshold $\delta J_\times(L)$. If $\delta J < \delta J_\times(L)$, $\Delta_{L/4}$ decreases more rapidly as $\delta J$ is reduced. The crossover between the two scaling regimes is well described by the data collapse shown in the inset, which utilizes a FSS form
\begin{align} \label{eq:DL4_fss}
	\Delta_{L/4} = L^{-1/2} \Psi_{\Delta}(\delta J^{1/2} L^{1/2})
\end{align}
with
\begin{align} \label{eq:psi_D}
	\Psi_{\Delta}(\zeta) = \begin{cases}
		\zeta &\text{ for $\zeta < 1$,}\\
		\zeta^2 &\text{ for $\zeta > 1$.}
	\end{cases}
\end{align}
This FSS form implies $\delta J_\times(L) = 1/L$. Combining this identity with $\epsilon_\times(L)$ obtained in Eq.~\eqref{eq:epsilon_crossover}, one finds $\delta J_\times(L) = e^{-b/\epsilon_\times(L)}$, which means that the two crossover scales are related to each other by Eq.~\eqref{eq:DJ-exp-fit}. Thus the linear (quadratic) regime of $\Psi_\Delta$ exactly corresponds to the strong (weak) SB regime.

In the strong SB regime, $\Delta_{L/4} = \delta J^{1/2}$ is a natural consequence of the nonzero excess bulk density $\Delta_\mathrm{b} = \delta J^{1/2}$ dominating most of the system, so that $\Delta_{L/4} = \Delta_\mathrm{b}$. On the other hand, in the weak SB regime, Eqs.~\eqref{eq:DL4_fss} and \eqref{eq:psi_D} imply
\begin{align} \label{eq:DL4_scaling}
	\Delta_{L/4} = \delta J \, L^{1/2},
\end{align}
which cannot be explained by the bulk effect. In order to understand this scaling behavior, we consider the relation between $\Delta_{L/4}$ and $\Delta_\mathrm{SB}$, where the latter denotes the excess particle density right next to the slow bond. According to our argument for $\epsilon_\times(L)$ discussed above, the weak SB regime is equivalent to the maximal-current phase, in which any density modulation algebraically decays to $\rho_\mathrm{b} = 1/2$ as a square root of the distance. Thus we have $\Delta_{L/4} \sim \Delta_\mathrm{SB}/L^{1/2}$ in this regime. For very small $\epsilon$, from $\Delta_\mathrm{SB} \sim \epsilon$ and Eq.~\eqref{eq:DJ-lin-fit}, we can derive Eq.~\eqref{eq:DL4_scaling}. Finally, since the weak SB regime is effectively a maximal-current phase without any special length scale, the entire regime becomes self-affine. Thus Eq.~\eqref{eq:DL4_scaling} should hold for the whole weak SB regime, which explains the sharp transition between the two different scaling behaviors in the inset of Fig.~\ref{fig:boundary_effects}.

There are notable similarities and differences between the weak SB regime described above and the nonqueued SB phase hypothesized in \cite{MHa2003}. Both are characterized by the power-law density modulations created by the slow bond, which are more dominant than the excess bulk density $\Delta_\mathrm{b}$. However, whereas the power-law density modulations of the weak SB regime seem to be characterized by an exponent $1/2$, the corresponding exponent in the nonqueued SB phase is $1/3$. The most importance difference lies in the fact that the weak SB regime vanishes as $L$ goes to infinity (thus it is not a ``phase''), while the nonqueued SB phase should remain.

It is natural to ask why our study and \cite{MHa2003} lead to such different scenarios, despite close similarities in the system setups and observables. We do not understand yet what causes the first difference mentioned above, but the origin of the second difference will be clarified in the next part of our discussion.

\begin{figure*}[]
	\centering
	\includegraphics[width=0.85\textwidth]{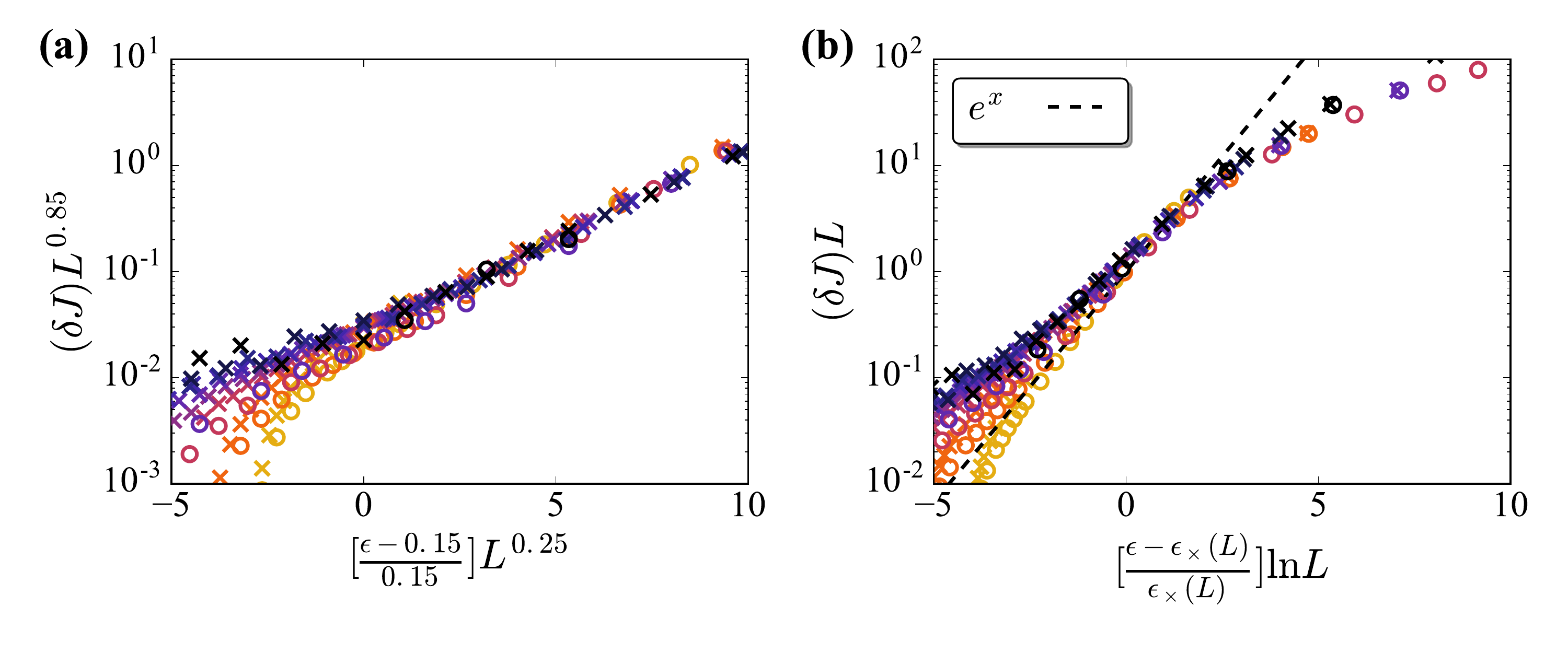}
	\caption{\label{fig:nonzero_ec_collapse} Two conflicting FSS theories for the OBC ($\circ$) and the PBC ($\times$). (a) Data collapse for Eq.~\eqref{eq:DJ_fss_nonzero_ec}, which corresponds to $\epsilon_\mathrm{c} = 0.15$. (b) Data collapse for Eq.~\eqref{eq:DJ_fss_alt}, which implies $\epsilon_\mathrm{c} = 0$. The system size is varied from $L = 2^6$ to $2^{16}$, with darker colors used for larger $L$  (see the legends of Fig.~\ref{fig:current}).}
\end{figure*}

\subsection{FSS forms for zero and nonzero $\epsilon_\mathrm{c}$}

All the above results claim that our numerical data are in good agreement with $\epsilon_\mathrm{c} = 0$. However, as shown in Fig.~\ref{fig:nonzero_ec_collapse}(a), the same data are also in good agreement with a different FSS form~\cite{MHa2003}
\begin{align} \label{eq:DJ_fss_nonzero_ec}
	\delta J = L^{-2\beta/\nu} \Phi_>\left[\frac{\epsilon - \epsilon_\mathrm{c}}{\epsilon_\mathrm{c}}\,L^{1/\nu}\right].
\end{align}
The unknown parameters of this FSS form are estimated to be $\beta \simeq 1.7$, $\nu \simeq 4.0$, and $\epsilon_\mathrm{c} \simeq 0.15$, whose values are comparable to those reported in \cite{MHa2003} ($\beta \simeq 1.46$, $\nu = 3.95$, and $\epsilon_\mathrm{c} = 0.20$). Here the nonzero value of $\epsilon_\mathrm{c}$ raises the question of why the same data apparently support both contradictory claims.

In order to address the question, we show that there is yet another FSS form, which implies $\epsilon_\mathrm{c} = 0$ but has structural similarities with Eq.~\eqref{eq:DJ_fss_nonzero_ec}. Since $\xi(J)$ and $L$ are two competing length scales giving rise to finite-size effects, we may construct a scaling function whose argument is given by
\begin{align}
	\ln \frac{\xi(J)}{L} \sim \frac{b}{\epsilon} - \ln L \sim \frac{\epsilon-\epsilon_\times(L)}{\epsilon_\times(L)}\,\ln L \nonumber
\end{align}
for $\epsilon$ in the vicinity of $\epsilon_\times(L)$. Then one can check that
\begin{align} \label{eq:DJ_fss_alt}
	\delta J = L^{-1} \, \Phi_0\left[\frac{\epsilon-\epsilon_\times(L)}{\epsilon_\times(L)}\,\ln L\right]
\end{align}
where $\Phi_0(\zeta) = e^\zeta$ for $\zeta > 0$ is consistent with Eq.~\eqref{eq:DJ-exp-fit}. As shown in Fig.~\ref{fig:nonzero_ec_collapse}, our data are in good agreement with this FSS form as well.

The data are well described by both Eqs.~\eqref{eq:DJ_fss_nonzero_ec} and \eqref{eq:DJ_fss_alt} because both FSS forms have components which are numerically hard to distinguish from each other. Namely, $\epsilon_\times(L) \sim 1/\ln L$, $\ln L$, and $L^{-1}$ in Eq.~\eqref{eq:DJ_fss_alt} are not easily distinguishable from $\epsilon_\mathrm{c} > 0$, $L^{1/\nu}$, and $L^{-2\beta/\nu}$ in Eq.~\eqref{eq:DJ_fss_nonzero_ec}, respectively, unless one can obtain the data for a very broad range of $L$. Therefore our FSS analyses {\it per se} are not enough to rule out the possibility of $\epsilon_\mathrm{c} > 0$. In spite of this limitation, our analyses clarify how the numerical results, previously thought to support $\epsilon_\mathrm{c} > 0$, can be reconciled with the analytical results supporting $\epsilon_\mathrm{c} = 0$ by accepting the hypothesis of a crossover between the strong SB regime and the weak SB regime. \newline

\section{Summary}
\label{sec:summary}

We revisited the controversial slow-bond (SB) problem via finite-size scaling (FSS) analyses. Building upon the previous conjectures about an essential singularity at the vanishing SB strength $\epsilon_\mathrm{c} = 0$~\cite{Janowsky1994,Costin2012}, we proposed FSS forms which arise from the competition between the correlation length and the system size $L$. The FSS form implies that the boundary between the strong SB regime and the weak SB regime is not a true dynamical phase transition hypothesized in \cite{MHa2003}, but a crossover SB strength which asymptotically vanishes as $\epsilon_\times(L) \sim 1/\ln L$. Our FSS theory was found to be in good agreement with the numerical data.  In view of a recent rigorous study~\cite{Basu2014} supporting $\epsilon_\mathrm{c} = 0$, it seems plausible that the FSS theory of \cite{MHa2003} should be replaced with the one proposed in this study.

We also found that the weak SB regime shares characteristics of the maximal-current phase with the nonqueued SB phase conjectured in \cite{MHa2003}. Still, except for the extreme case when the SB strength $\epsilon$ is small enough to make each blocking event independent, the properties of this regime remain largely unknown. Many poorly understood questions, such as how the scaling behaviors change for larger $\epsilon$, and whether the exponent $1/3$ of the density modulations identified in \cite{MHa2003} has any place in our FSS theory, are left as the subjects of future studies.

\acknowledgements
{This work was supported by the National Research Foundation of Korea (NRF) funded by the Korean Government [Grants No. NRF-2014R1A1A4A01003864 (H.S., M.H.) and No. NRF-2015-S1A3A-2046742 (H.S., H.J.)]. Y.B. was supported in part at the Technion by a fellowship from the Lady Davis Foundation.}
\bibliography{ref-EX11323-final}

\end{document}